\documentclass[11pt]{article} \usepackage{hyperref} \pdfoutput=1
\begin{document} 

\title{Vortex Rings from {\em Sphagnum} Moss Capsules}

\author{Emily S. Chang$^1$, Joan Edwards$^2$, Jung Han Cha$^{1}$, Sam Strassman$^{1}$, \\Clara Hard$^2$, and Dwight L. Whitaker$^1$ \\ \\ $^1$ Department of Physics and Astronomy, \vspace{6pt} \\Pomona College, Claremont, CA 91711, USA,\\ \\$^2$Department of Biology, \vspace{6pt}\\Williams College, Williamstown, MA 01267, USA}

\maketitle
%% The abstract (in this file, and that submitted as text to arXiv) should include the exact phrase %% "fluid dynamics video" or "fluid dynamics videos"
\begin{abstract}   Long distance wind dispersal requires small spores with low terminal velocities, which can be held aloft by turbulent air currents until they are deposited in suitable habitats for colonization. The inherent difficulty in dispersing spores by wind is that spores easily carried by wind are also rapidly decelerated when moving through still air. Thus the height of spore release is critical in determining their range of dispersal. Vascular plants with wind dispersed spores use the height of the plant to lift spores into sufficient wind currents for dispersal, however non-vascular plants such as Sphagnum cannot grow sufficiently tall. These fluid dynamics videos show how exploding capsules of {\em Sphagnum} moss generate vortex rings to carry spores to heights above 10~cm with an initial velocity of 16 m\,s$^{-1}$.   In contrast spores launched ballistically at these speeds through still air would travel only 2-7 mm. 
 \end{abstract}

% main text \section{Introduction}

{\em Sphagnum} or peat moss is an abundant genus of plants that covers over 1\% of the land are on earth and stores more carbon than any other genus.  There are over 285 species of {\em Sphagnum}, often with specific microhabitat
requirements based on pH and water level, and suitable habitats (e.g., kettle bogs) are often
separated by many kilometers. Most colonization of new and distant habitats is by spores, thus
long distance dispersal along with prolific spore production are critical in maintaining this genus.  This paper describes videos showing the formation of vortex rings to effect the long range spore dispersal.  

{\em Spahgnum} capsules were collected from a kettle bowl bog in Pownal, VT during the summers of 2005-2009 and brought to laboratories at Williams and Pomona Colleges to be recorded with high-speed digital cameras.  Capsules change shape from spherical to cylindrical when they are mature, which causes pressure to build up inside as air is compressed.  These shape changes typically happen over 5-10 minutes and cylindrical capsules will typically explode within a minute under the bright lights used for recording videos.  

The first clip in our video shows a distant shot of spore ejection recorded at 1,000 frames\,s$^{-1}$ in which you can see that spores emerge in a tight stream from the capsule that slows gradually.  The next clip shows a close-up of the capsule explosion recorded at 11,000 fps with an exposure time of 20 $\mu$s.  Here the mushroom cloud from the formation of a vortex ring is clearly visible.  The spores are light enough that most of them remain entrained in the vortex bubble throughout this clip, however the spores are too massive to track the circulating fluid in the vortex ring.  

The last two clips are of computer simulations of the capsule explosion.  These simulations have assumed an internal pressure of 3 atm in the capsule and were performed with a large eddy simulation (LES) using ANSYS FLUENT 12.  In the first simulation video the sudden apparent bursts of speed are a result of a change in time steps in the simulation and do not indicate an acceleration of the vortex bubble.

This video has been submitted to the American Physical Society Division of
Fluid Dynamics annual meeting showcase, the Gallery of Fluid Motion 2010 and can be seen at these links:

\begin{itemize}

\item High resolution  \href{/Whitaker_Sphagnum_hires}{Video 1} 

\item  Low resolution \href{/Whitaker_Sphagnum_lowres}{Video 2 } 

\end{itemize}

\end{document}